\documentclass[aps,pre,reprint]{revtex4-1}
\usepackage{graphicx}
\usepackage{dcolumn}
\usepackage{bm}
\usepackage{hyperref}

\draft 

\begin{document}




\title{Breakdown of metastable political duopoly due to asymmetry of emotions in partisan propaganda} 
\author{Pawel Sobkowicz}
\email[]{pawelsobko@gmail.com}
\affiliation{KEN 94/140, 02-777 Warsaw, Poland}


\date{\today}

\begin{abstract}
We present results of opinion dynamics simulations based on the emotion/information/opinion (E/I/O) model, applied to
a strongly polarized society. Under certain conditions the model leads to metastable coexistence of
two subcommunities (supporting each of the opinions) of comparable size -- which corresponds to bipartisan split found in many real world communities.
Spurred by the recent breakdown of such system, which existed in Poland for over 9 years, we extend the model by allowing a third opinion.
We show that if the propaganda messages of the two incumbent parties differ in emotional tone, the system may be ``invaded" by a newcomer third party
very quickly -- in qualitative agreement with the actual political situation in Poland in 2015. \end{abstract}

\pacs{89.65.-s, 89.75.Fb, 02.70.-c}


\maketitle 

\section{Introduction}
\label{sec:introduction}

Studies of opinion changes in societies are part of the core of topics of sociophysics.
One of the reasons is the importance of understanding of changes in public attitudes versus specific issues or policies.
There are many approaches, differing in the description of the available opinion states, interpersonal dynamics, network description and many others. Among the most popular, one can mention the voter model
\cite{cox86-1,bennaim96-1,galam02-4,castellano03-1}, the Sznajd model
\cite{sznajd00-1,stauffer01-2,stauffer02-1,stauffer02-2,slanina04-1,sabatelli03-1,sabatelli03-2,bernardes01-1},
the bounded confidence model \cite{deffuant00-1,deffuant02-1,weisbuch03-1,weisbuch03-2,sousa04-1,lorenz07-1},
the Hegelsmann-Krause model \cite{hegelsmann02-1}, the social impact model of Nowak-Latan\'{e} \cite{nowak90-1,nowak96-1} and
its further modifications \cite{holyst01-1,kacperski99-1,kacperski00-1,sobkowicz10-2}.

In a series of previously published papers \cite{sobkowicz12-7,sobkowicz13-2,sobkowicz13-3} we have introduced a model that combines the dynamics of individual opinion and emotion changes.
In the current work we use a modified version of the model (introducing a possibility of three opinions, corresponding to three political parties) to attempt a description of the recent changes on the Polish political scene.  

The paper is organized as follows: Section II describes shortly the Polish political situation. Section III describes shortly the model while Section IV focuses on the introduced modifications. Section V describes the results of the model and the dependence on the parameters. Section VI provides a discussion of the results and possible developments of the situation for the upcoming parliamentary elections and afterwards.

\section{Political situation in Poland}
\label{sec:political}

During the 25 years since the overthrow of the communist rule, Poland has had a quite active political history. With a single exception in 2011, every parliamentary election brought a radical change in the landscape of political parties and the ruling coalition. The  election rules were changed profoundly: initially they allowed people to vote for candidates representing multiple parties; later they changed to votes for one party, in a proportional system with rather high threshold of 5\%. This later change was coupled with a growing polarization of the political scene (and the whole society), which has led eventually to a dominance of two parties: Platforma Obywatelska (PO, Civic Platform) and Prawo i Sprawiedliwo\'s\'c (PiS, Law and Justice). It is rather interesting, that in the parliamentary elections in 2005 which brought the two parties to the dominant positions at the expense of the leftist parties, the election campaign was based on the promise of a joint government, with the unofficial acronym POPiS. 
This was considered by many voters to be a very attractive political idea, and subsequently, the two parties got 28.91\% and 33.70\% of the votes, respectively.
No other party received more than 13\% of the votes. 
Yet shortly after the elections a conflict between the leaders of the two parties broke out, and the promised coalition never happened, leaving a part of the electorate dubbed ``POPiS orphans".
The government was formed by PiS, led by Mr Jaros{\l}aw Kaczy\'nski, in coalition with two smaller parties. 
This proved untenable, and in early elections in 2007 due to the crisis, it was PO that captured  the majority (41.51\%), while PiS has became a strong opposition, with 32.11\% of votes. 
This result was repeated in the next parliamentary elections in 2011, which was won by PO (39.18\%) with PiS receiving 29.89\% of the votes.

Since 2005 until 2014, the two parties have enjoyed a virtual dominance on the Polish political scene, as documented not only by the results of the parliamentary elections but also the presidential ones, elections to the European Parliament and local elections. It is also confirmed by the data from popularity polls, which show practically stable, high level support for the two dominant parties with a stable or decreasing popularity of the minor parties. The stability is even more visible if one adds to the results of PiS the figures for two smaller parties: PJN and SP, which have split off from PiS in 2011 and 2012, due to personal differences, keeping a very similar political program and addressing the same electoral base. The two parties merged back with PiS in 2014. 

\begin{figure}
\includegraphics[scale=0.3]{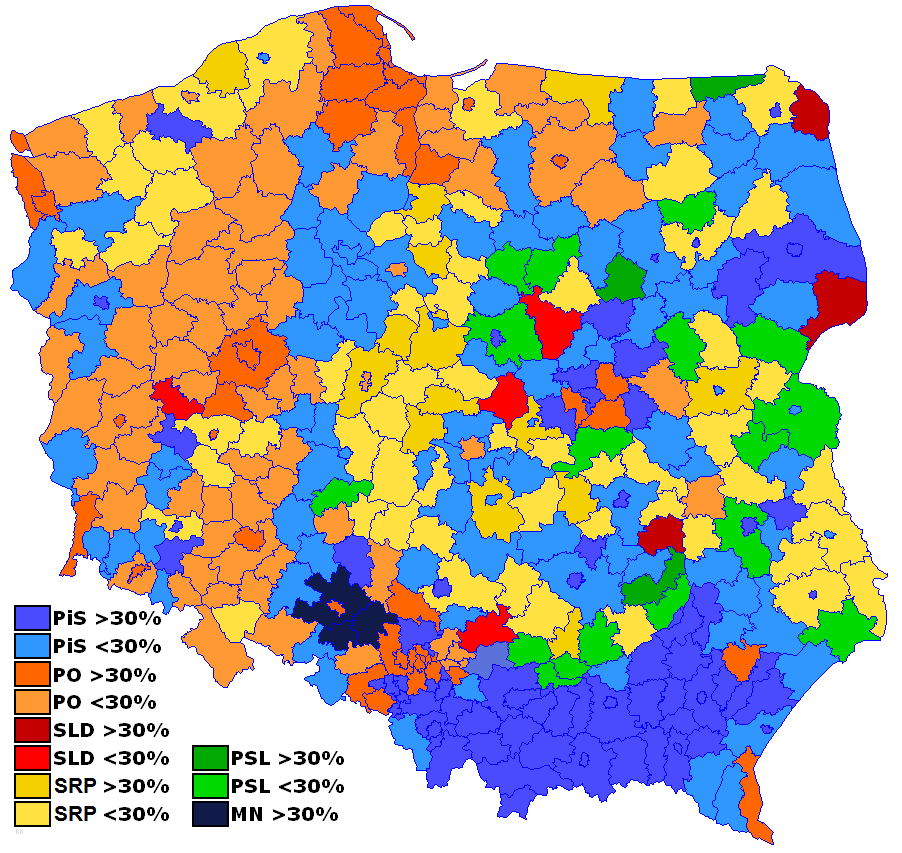}
\includegraphics[scale=0.23]{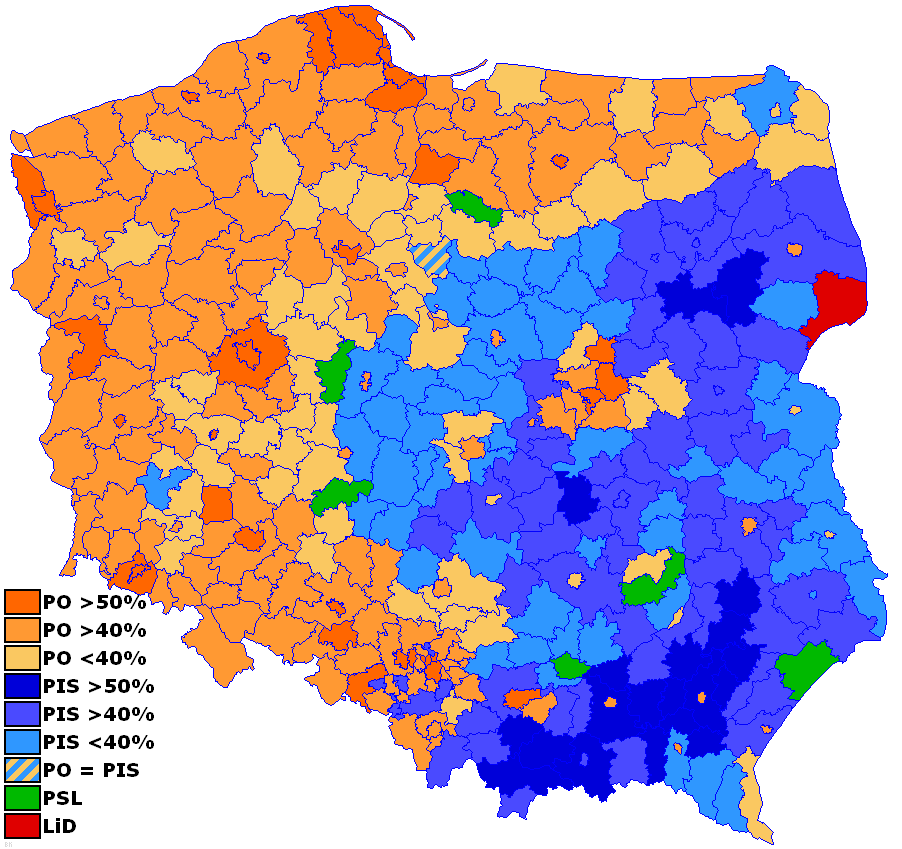}
\includegraphics[scale=0.3]{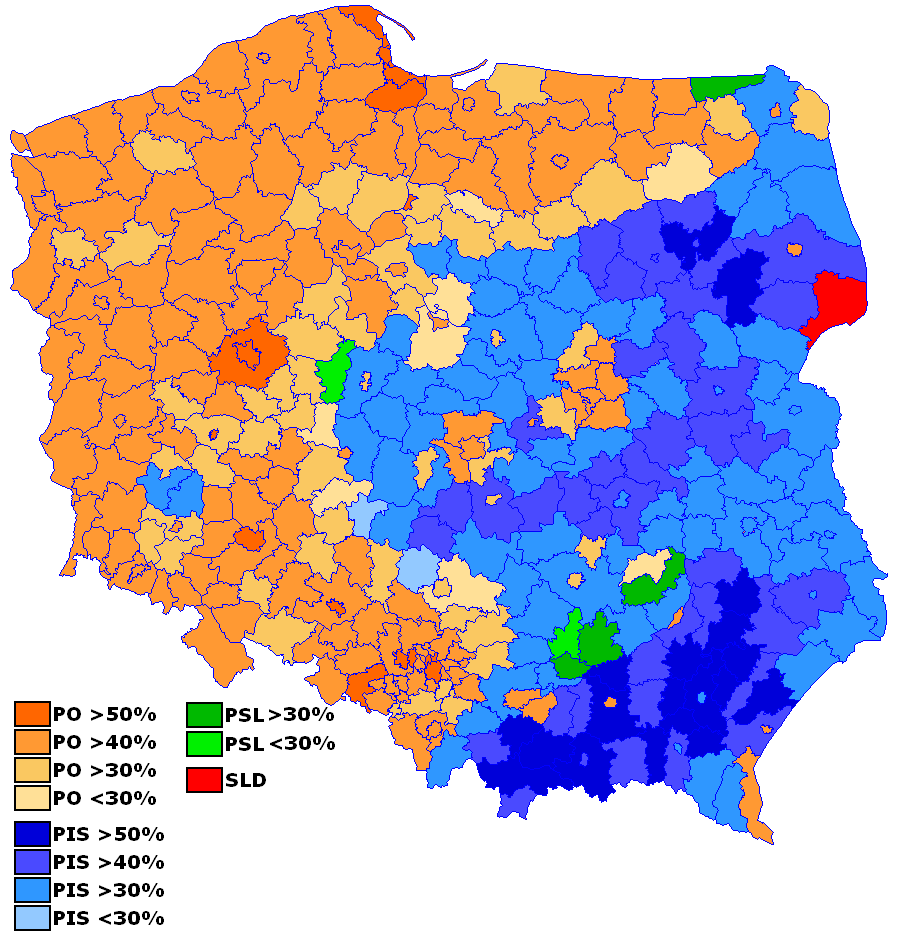}
\caption{Evolution of the local community voter preferences in Polish parliamentary elections in 2005 (top), 2007 (middle) and 2011 (bottom). Source:  Robert Wielg\'orski (Barry Kent) to be found under \url{https://pl.wikipedia.org/wiki/Wybory_parlamentarne_w_Polsce}.
\label{fig.geography}}
\end{figure}

There are multiple explanations for the duopoly that emerged in Poland in 2005. On one level, it is quite illuminating to consider the map of voter preferences, which shows remarkably stable pattern of support, roughly corresponding to historical boundaries of Polish partitions in 19th century. Such geographical division of preferences corresponds to deeper cultural patterns, which will not be discussed in this work. We note that the changes in the voting results take the form of gradual shifts of the dominance in the constituencies, as shown in Figure~\ref{fig.geography}. This pattern justifies, to a certain extent, the use of a simple two-dimensional topology for the agent based model.

Another reason for the stability, in our opinion the crucial one, is the capturing of the public debate by the two parties.  Ever since the elections of 2005, the would-be coalition partners started an aggressive fight, leading to a very strong polarization of the Polish population and media. Much of the emotional content was focused on personal differences between the leaders of the two parties: Mr jaros{\l}aw Kaczy\'nski and Mr Donald Tusk, the leader of PO. Such polarization has left very little room for any other political option/party -- phenomenon known in other countries (e.g. the Democrat/Republican split in the US political scene). The public campaigns of both parties focused on negative emotions, accusations, fear of what would happen if `they' win, etc.
This aggressiveness has increased, especially on the part of PiS , since the crash of the plane carrying the Polish President, Mr Lech Kaczy\'nski (twin brother of the PiS leader) near Smolensk in Russia in April 2010.  In the subsequent  presidential elections of 2010, Mr Jaros{\l}aw Kaczy\'nski (who stepped in to continue his brother legacy) lost to the PO candidate, Mr Bronis{\l}aw Komorowski. In this way PO has achieved  full control of the government structures: parliament majority with a single coalition partner, government and the presidential office (but short of of reaching the capacity to change the constitution).

The tragic plane crash and the lost elections have significantly increased the negative tone of the PiS communications with the electorate. In addition to the typical (for a political opposition) criticism of mis-management of the country, a significant part of the message focused on accusations that the plane crash was, in fact, a result of some sort of a plot, and that the plane was destroyed by a mid-air explosion. Implicitly, the PiS propaganda indicated Russia as the culprit, but the direct target was the PO government, either as acting in collusion with the perpetrators of the attack, or, in the least aggressive case, as grossly incompetent in handling of the crash investigation.

\begin{figure*}
\includegraphics[scale=0.55]{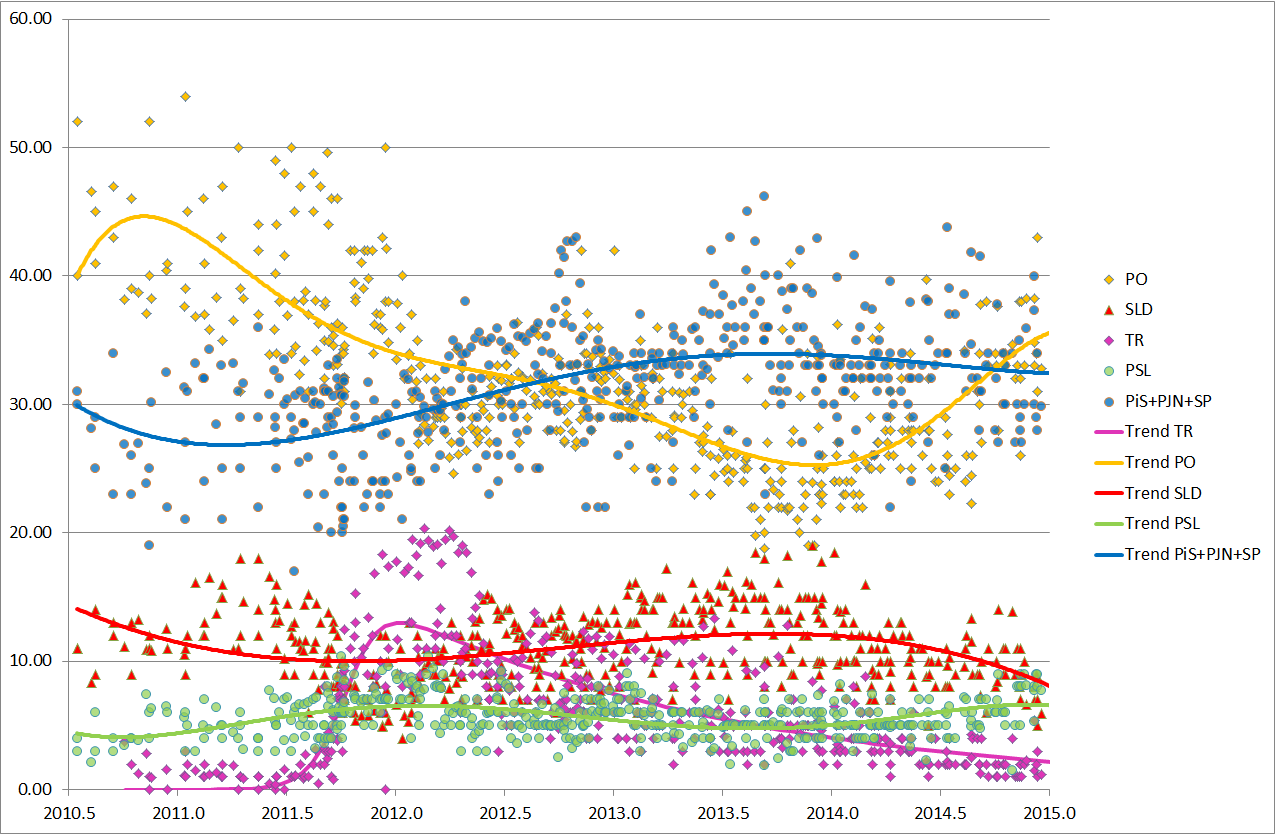}
\caption{Evolution of the support for major political parties in Poland 2010--2014. PO and PiS are the main contenders, while SLD (socialdemocrats) and PSL (peasants party) are smaller ones, existing on the scene during the whole 25 year post-communist era, with small but well entrenched core electoral base. The data on PiS group them together with two other parties, PJN and SP, which have split-off from PiS during the period, but which have since returned to form a single political entity in late 2014. As the three parties address the same electorate with very similar propositions, we treat them together in the polls analysis.
Since 2007, PSL is the coalition partner of PO. 
TR (Ruch Palikota/Twoj Ruch) is a party formed in mid-2011 by a PO dissident, which has enjoyed a brief period of success between 2011 and 2012.  Data from Mr Maciej Witkowiak, \url{http://niepewnesondaze.blogspot.com/}. \label{fig:parties} }
\end{figure*}

\begin{figure}
\includegraphics[scale=0.55]{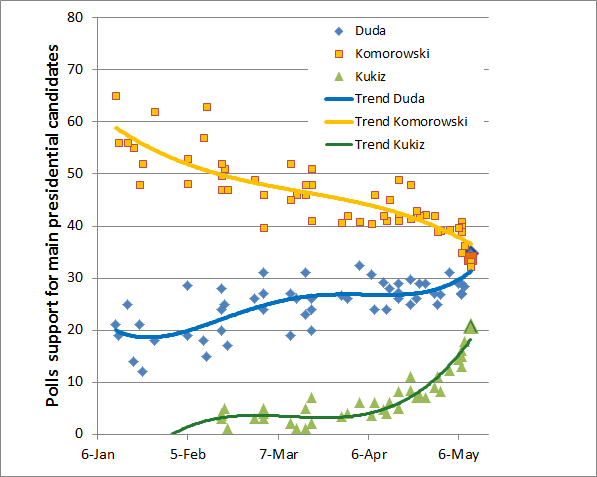}
\caption{Evolution of the support for presidential candidates of PO (Mr Komorowski) and PiS (Mr Duda) in the 2015 presidential elections, including the support for the independent candidate, Mr Kukiz. The final (larger) points indicate the results of the first round of voting in the presidential elections. Data from various polls, via Wikipedia. \label{fig:president} }
\end{figure}

\begin{figure*}
\includegraphics[scale=0.55]{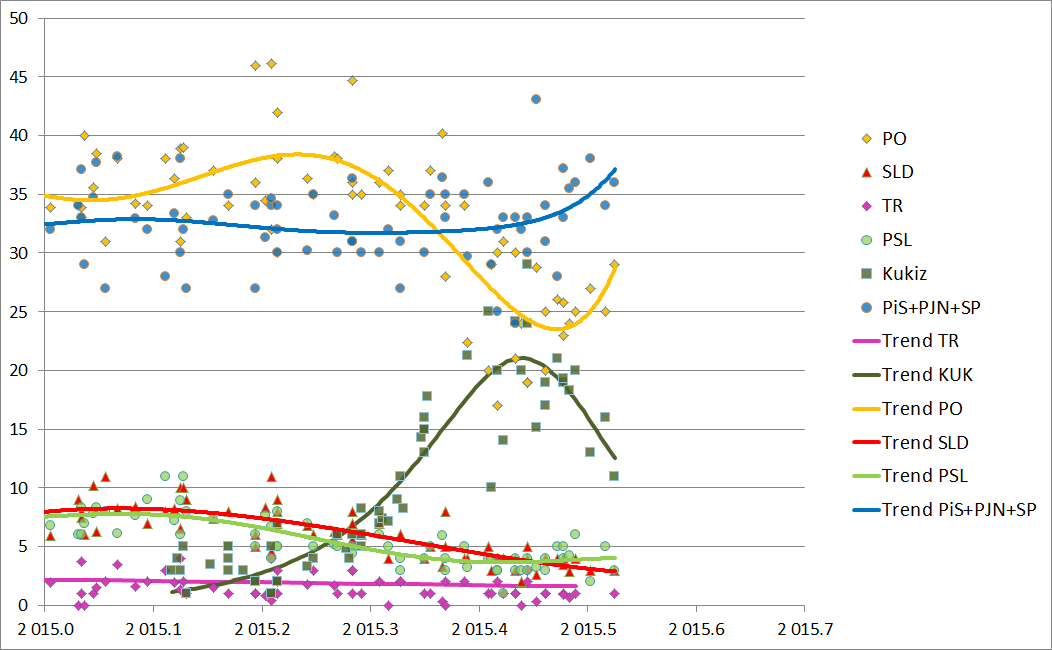}
\caption{Evolution of the support for major political parties in Poland in 2015. The data for Mr Kukiz party (which started to get recognized in the polls since June 2015) is spliced with his personal support in the presidential elections, shown in Figure~\ref{fig:president}.   Due to the nature of the movement, based largely on the personality of the leader, we think such data use is permissible. Data from Mr Maciej Witkowiak, \url{http://niepewnesondaze.blogspot.com/}.  \label{fig:parties2015} }
\end{figure*}

The increasing conflict and the personal attacks between politicians representing the two camps resulted in an increasingly polarized society. Moreover, the polarization covers also the media outlets: there are TV stations, daily journals, weeklies and WEB portals that cover  almost exclusively a single viewpoint.
There is strong tendency to limit the range of the media used by the supporters of each political camp to the outlets representing the same views, and to describe the other channels as traitorous or trash.
  
The aggressively  negative communications
by PiS is in stark contrast with the communication strategy chosen by PO. For example, the government has been very late in mounting an information campaign countering the conspiracy theories of the Smolensk crash embraced by PiS. Relying on the results of the official investigations, the messages were long delayed, and lacked the emotional appeal.
At the same time, shortly after the presidential elections in 2010, Mr Tusk declared that ``as long as he is active in public life, he prefers, what some mischievous commentators call warm water in the tap policy"
\footnote{Gazeta Wyborcza, 20 Sept 2010, \protect\url{http://wyborcza.pl/1,76842,8397290,Tusk__Wole_polityke_cieplej_wody_w_kranie.html}}.
Most of the communication strategy of PO focused on the achievements of the government and country as a whole, its economic growth, new infrastructure such as highways, etc.
It should be noted here, that during the period in which PO was in power, Poland has indeed enjoyed significant benefits of the EU membership and has weathered the economic crisis with remarkable success, being the only country to preserve continuous growth  of GDP (dubbed ``green island" by the PO government). Thus the ``success" focused communications did have
some real background. But they did not evoke emotional interest and commitment on the part of the PO supporters comparable to the effects of the negative campaign of PiS.
Following the election of Mr Tusk to the post of the President of the European Council in late 2014, PO has focused even more its media campaign on the positive image ``all is well" messages, clearly ignoring the increasing signs of dissatisfaction and demobilization of their own electoral base.
Even so, the local elections held in November 2014 resulted in a tie between PO and PiS, with a very strong result for the PO coalition partner PSL.

Again, the reaction to these results was very different in the government camp (which has declared yet another victory), and PiS, which openly claimed the elections to be manipulated and the results rigged. It is no surprise that the emotional response in the aftermath of the elections was very different as well: while the PO voters mostly did not take much notice, the PiS electorate was actively mobilized and formed many ``spontaneous" initiatives ``to protect future elections".

The long period of relative stability of the PO-PiS duopoly described above seems to be broken by an emergence of a third power with comparable political support, as shown by the events related to the recent presidential elections held in May-June 2015. The initial polls, in January 2015, have indicated a landslide victory of the incumbent president, Mr Komorowski (with over 60\% support, allowing to win in the first round of voting). The expectation of an  easy win was further enhanced by the choice of a  candidate by PiS (instead of the party leader, Mr Jaros{\l}aw Kaczy\'nski, the party has chosen as its candidate Mr Andrzej Duda, almost unknown to the general public). A lackluster campaign by the incumbent has led to a steady decrease of the support, while the support for Mr Duda has grown significantly. Eventually Mr Duda won the first round (gathering 34.76\% of the votes, compared to 33.77\% for Mr Komorowski) and the second round (albeit by a rather narrow margin 51.55/48.45\%). At the same time, the polls conducted in May and June 2015, ahead of the parliamentary elections (to be held in late 2015) show a significant and increasing lead of PiS over PO, reversing the order of the support present in the past decade, mostly due to the dramatic drop in the PO support.
It should be noted here, that the external campaign of Mr Duda was much more toned down and focused on his image as a rational, future oriented politician, rather than the typical PiS aggressive messages.

What is the most interesting phenomenon, which has prompted this particular work, was an apparent breakdown of the political duopoly. 
An independent candidate, Mr Pawe{\l} Kukiz (rockman, who previosly has not been much active in politics). His campaign, which started almost two months after the main rivals', lacked significant funding, but has focused the attention of the dissatisfied, especially the young part of the population. Based on a single battle cry for single-member constituencies (such a change from the current proportional system would require changes in the Polish constitution), Mr Kukiz appealed to all the dissatisfied with the current party  system. 
In the space of a few months he has gathered  an enormous support, getting 20.8\% of the votes in the first round of the elections (see Figure~\ref{fig:president}). 
After the elections, a wider movement coalesced around Mr Kukiz, becoming the third political power as measured by the polls in early summer 2015. Mr Kukiz continues to push for the single-member constituencies and the parliamentary elections, scheduled for October 2015.

Achieving popularity at the level of two thirds of the support for the mainstream candidates by an unknown candidate is a phenomenon that can not only change the future of Polish politics (which remains to be seen) but also requires an in-depth scientific study. While stable duopolies have already been studied via agent based models, the sudden breakdown of such situation provides and interesting challenge.
In our work we aim to provide a simple agent based model, which shows, in a single conceptual frame, both the previous stable duopoly and its recent breakdown, along with some possible future scenarios.

\section{General model description}
\label{sec:model}

Our goal is to present an agent based model that would be simple enough to understand intuitively, yet which would be based more closely on psychological understanding of human behavior than the standard approaches. Part of the motivation comes from extended studies of the behavior of the users of Internet fora related to the Polish politics between 2009 and 2011, where we have observed strong correlations between the expressed opinions and emotions of the participants \cite{sobkowicz2010dynamics,sobkowicz12-2,sobkowicz12-4}. Notably, we have observed that when the emotional arousal of the participants is high, their capacity to change opinions is negligibly small. This observation has led us to propose an approach in which the individual opinion about a specific issue would be influenced by a combination of the information related to the matter and the emotional state of the person.

The current paper is based on the model introduced in \cite{sobkowicz12-7} and \cite{sobkowicz13-2}, in which the model is described in considerable detail. The structure of the model combines  ``microscopic" description of individual opinion change with a flexible communication mechanism, allowing to use the model in different social contexts. The proposed solution is based on a simplification of the catastrophe theory of behavior, which has been introduced more than 30 years ago \cite{zeeman76-1}, and had been used (and criticized) in  analyses of human
behavior (for a review see \cite{rosser07-1}).
The cusp catastrophe model allows to describe situations where the same amount of information that would lead to an opinion change for low emotional state would leave the agent's opinion unchanged for high value of the emotional arousal.
While the catastrophe theory has been used in description of the individual behavior,  it is not well suited for large scale opinion simulations. This is because the continuous nature of the control variables makes it very difficult to correctly map the model and psychological observations and then to assign these values to computer based agent societies -- there is simply too much variability in the starting conditions and system evolution.

For this reason we have proposed \cite{sobkowicz12-7,sobkowicz13-2} a discrete version of the approach, in which the continuous folded cusp surface is replaced by just seven states corresponding to two values of the emotion level (splitting factor): calm and agitated and three values for the information and opinion: pro, contra and neutral.
This can be easily transformed to descriptions of support for political parties, especially in a duopoly situation.

We are using an approach based on communication via discrete messages. Such message -- originating from another agent or from the press, TV, the Internet or other media -- would be described by exactly the same set of variables as the recipient agent: emotional arousal level, information and opinion. In the case of messages sent by an agent, the characteristics of the message are assumed to equal those of the authoring agent. 
The use of discrete states allows a simple description of  behavior of an individual agent resulting from contact with another member of the society or with an external information/emotion source.

Upon receiving a message (which means either interacting with another agent or with media carrying some propaganda messages), the recipient state may, in some cases, be modified. Generally, the model assumes that agents in calm states are capable of changing their opinion, if they receive information contrary to their current beliefs (calm stale allows rational processing of such information leading to change of the opinion). On the other hand, agitated agents, even exposed to contrary information, would preserve their opinion unchanged, refusing to accept /process the messages in a rational way. Thus, within the model, the only way to make an agitated agent change its opinion is first via calming it down (which may happen if the agent is in contact with calm messages or agents with which it is in agreement), and only then by changing the opinion by contact with messages or agents with an opposing view.
In addition to the ``calming" processes, a reverse process in which a calm agent exposed to contrary view may become agitated (without changing the opinion). This happens with a probability $p_{agit}$ when the contact is with a calm opponent/message and with a 100\% probability if the opponent is agitated.

These basic rules correspond to simplified, intuitive psychological description of opinion change of a person, dependent on the current opinion and emotional state of that person, as well as on the state of the person with whom the interaction occurs, again, both the opinion and the emotions.

As already mentioned in the current scenario we consider both direct interactions between agents and global messages. The latter are generated with specific opinions and end emotions (corresponding to the propaganda strategy of the political parties), issued with pre-fixed frequency. The reception of the agents to these messages depends on the  agreement or disagreement between the agent in question and the message, reflecting the phenomenon of selective attention known from social psychology \cite{stroud10-1,knobloch2012selective}.

The combination of the individual opinion and emotion dynamics and the message based communication algorithm has been already used to describe, in a quantitative way, the properties of a discussion forum, including the ratios of expressed emotions, opinions, types of messages (agreements, disagreements, trolling etc.) \cite{sobkowicz13-3}.   

\section{Specific simulation conditions}
\label{sec:simulations}

In the current approach we aimed at qualitative description of the real political situation. For this reason, in the following discussion, we will be using the names of the actual parties and political movements, rather than rely on abstract notation. This should facilitate the comparison between the model and the political situation described in Section~\ref{sec:political}, especially for non-Polish readers.  It should be noted that the use of this shorthand notation does not mean either support or lack of support on the part of the author for any of the parties mentioned.  

The two-opinion model used in the previous works may be easily extended to a larger number of exclusive opinions. In our case, considering three parties (two major contenders and a newcomer), instead of the seven states corresponding to the two party situation, there are 10 states: one global neutral and three states for each party. However, all the basic rules of the individual opinion dynamics remain the same.

\begin{figure}
\includegraphics[scale=0.4]{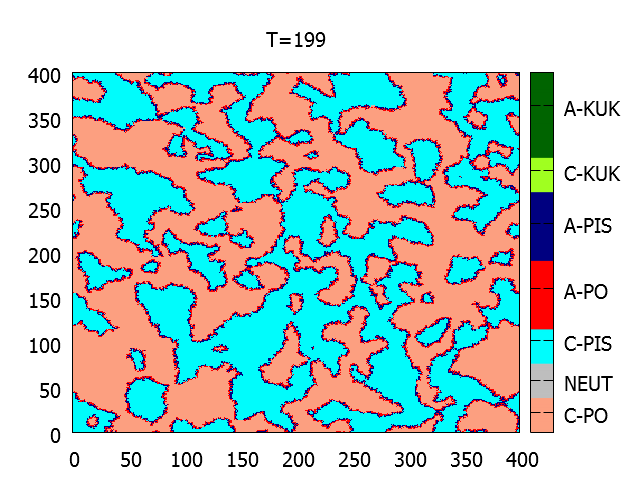} 
\caption{A snapshot of the system configuration at $T=199$  -- just before the PO and PIS propaganda messages are switched on. This time corresponds to the evolved starting conditions ($T_1$). Light blue: calm PiS supporters; dark blue: agitated PiS supporters; orange: calm PO supporters; red: agitated PO supporters. 
The spatial separation of the support base is reminiscent to the geographical 
patterns shown in Figure~\ref{fig.geography}.
Agitated agents are localized at the boundaries of the ``party held" domains.  \label{fig:frame199}}
\end{figure}

To describe the interactions between the agents we have chosen a simple 2D square geometry, as described in  \cite{sobkowicz13-2}. 
This allows us to visualize the evolution of the opinions. 
The system starts in a neutral state, with a small, randomly placed admixture of `seed' agents representing the two parties: PO and PiS. 
This admixture is at the level of $\sim$1\%. We have allowed only short range interactions between the agents, in the Moore neighborhood.   
In the absence of media messages, the system evolves to stable domains of shared opinions. 
Within these domains, agents are surrounded by neighbors sharing the same opinion, so their emotional state becomes calm. Agitation occurs at the boundaries between these domains. Such situation may be seen in Figure~\ref{fig:frame199}. This stage of the simulation ends at the time $T_1$, and may be considered a system preparation phase, in which the social division is created, so that the state at $T_1$ is a `true' starting condition. In the presented simulations $T_1=200$ Monte Carlo steps per agent.

Small modifications of the initial number of seeds representing PO and PiS lead to different ratios of supporters of the two parties at $T_1$. Up to this moment the system dynamics is fully symmetric between the two parties, and the final configuration at $T_1$ depends on the initial seed ratios.

\subsection{Treatment of party propaganda}

The treatment of the propaganda messaging in the system is the crucial development of the current work. These messages differ from the agent-to-agent messaging in several ways. 

First, the propaganda is divided into the categories of ``internal" (addressed to the party supporters) and ``external" (addressed outside the current support base. The internal propaganda may take two forms: ``mobilizing" (aimed at increasing the emotional commitment of the supporters, transitioning them from the calm into the agitated state); and ``demobilizing", which act in the opposite way: they make the agitated agents calm, and on top of that, they make the calm agents bored, turning them into the neutral state. 
As one can guess, this division is intended to mimic the main propaganda strategies of PiS and PO. 

The external propaganda comes again in two types: ``rational" -- aimed at converting neutral agents into the party supporters, and, additionally, converting calm supporters of other parties into neutral agents via rational argumentation. The latter effect may backfire, as with some probability $p_{agit}$ the agent receiving the message may become angered by it and turn from its calm state to an agitated one (just like the effects of an encounter between two calm agents supporting different parties). The value of $p_{agit}$ was set at 0.2 in the simulations.  The ``irrational" messages are, in a sense, a spillover from the internal mobilizing propaganda, and their effect is twofold: they change neutral agents directly to the agitated supporters, but they turn calm opponents into agitated opponents. 

The ratios of the various types of propaganda allow to model the actual behavior of he parties. Once the propaganda is switched on, that is after $T_1$,  the $P_{ratio}$ parameter determines the ratio of the propaganda messages to the total number of messages an agent receives on the average, in the simulations we have used $P_{ratio}=0.8$, i.e. 80\% of messages received by an agent resulted from the propaganda (his own party and other parties), while 20\% were results of the interactions with the closest neighbors. The ratios of various propaganda messages at this stage are presented in Table~\ref{tab:propaganda1}. 

\begin{figure}
\includegraphics[scale=0.4]{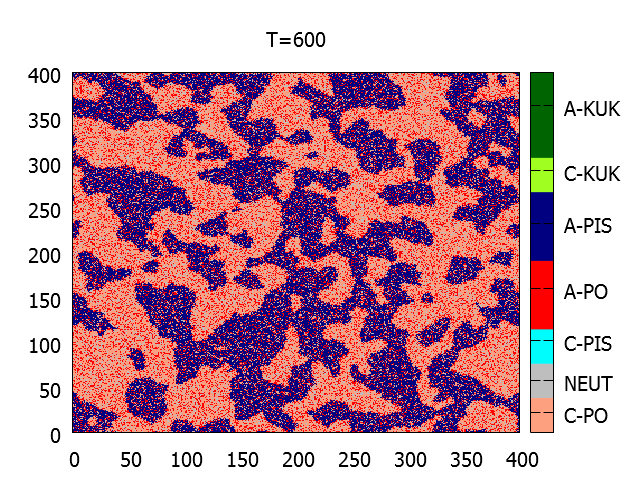} 
\caption{A snapshot of the system configuration at $T=600$ -- when the effects of the PO and PiS propaganda are well set in.  Almost all PiS agents are agitated, while PO domains remain relatively calm.  \label{fig:frame600}}
\end{figure}

At $T_1$ both major parties ``switch on" their propaganda machines. Until $T_2=800$ MC steps per agent, the political status quo is reproduced, with most of the PO supporters within their domains remaining calm, while most of the PiS supporters within their domains are turned into the agitated state. A typical snapshot of such situation is shown in Figure~\ref{fig:frame600}. 

Provided we limit ourselves to short range interactions between the agents, for a large range of the parameters the system quickly reaches a meta-stable  state, as noted in \cite{sobkowicz13-2}. This state would correspond to the long period of the duopoly between PO and PiS in the actual Polish politics, as shown in Figure~\ref{fig:parties}. 

\begin{figure}
\includegraphics[scale=0.4]{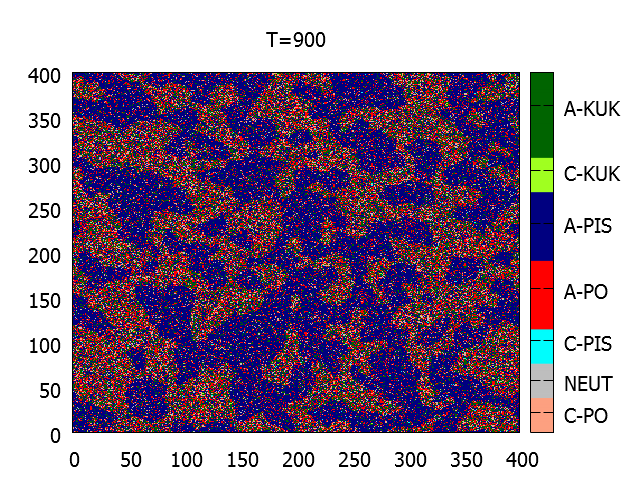} 
\caption{A snapshot of the system configuration at $T=1000$ -- 100 time steps after Kukiz propaganda started.  The agitated PiS agents are largely immune to it, while the previously calm PO domains become invaded by Kukiz supporters. Light green: calm Kukiz supporters; dark green: agitated Kukiz supporters.  \label{fig:frame900} }
\end{figure}

At $T_2$ the third party enters the political scene, with a mixture of internal and external propaganda messages. The parameters for this stage are shown in Table~\ref{tab:propaganda2}.  Thanks to the asymmetry in the system state just before $T_2$, the reception of the Kukiz-oriented propaganda is vastly different for the PiS and PO agents. It quickly gains support within the receptive PO dominated areas, while receiving practically no response in the PiS domains (Figure~\ref{fig:frame900}).  
In fact, at the same time PiS makes significant inroads into the areas dominated previously by PO, weakened by the Kukiz ``invasion". 

\begin{figure}
\includegraphics[scale=1.0]{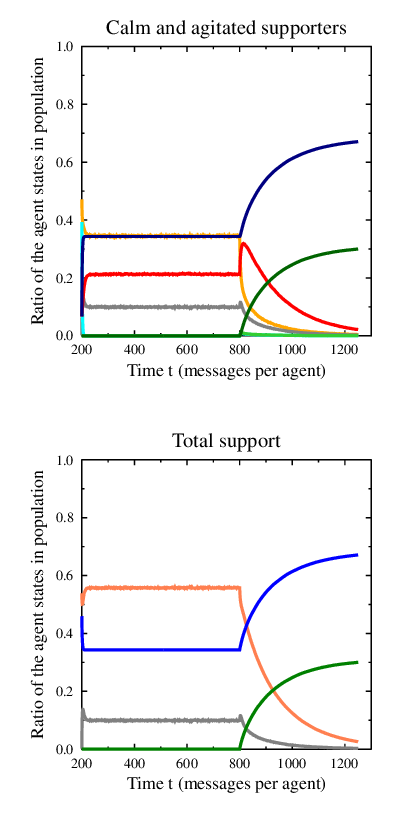} 
\caption{An example of the evolution of the support for the three political camps. Top panel: calm and agitated supporters; orange -- calm PO agents, red -- agitated PO agents; light blue -- calm PiS agents, dark blue -- agitated PiS agents; light green -- calm Kukiz agents, dark green -- agitated Kukiz agents; grey -- neutral agents. 
Bottom panel: evolution of the total number of supporters for each party. Orange -- PO, dark blue -- PiS, green -- Kukiz. \label{fig:evolution2}}
\end{figure}

Eventually, only a small number of agitated PO supporters remain, their initial numbers `eaten' from within by Kukiz supporters and from outside by PiS ones (Figure~\ref{fig:evolution2} 

\begin{figure}
\includegraphics[scale=0.4]{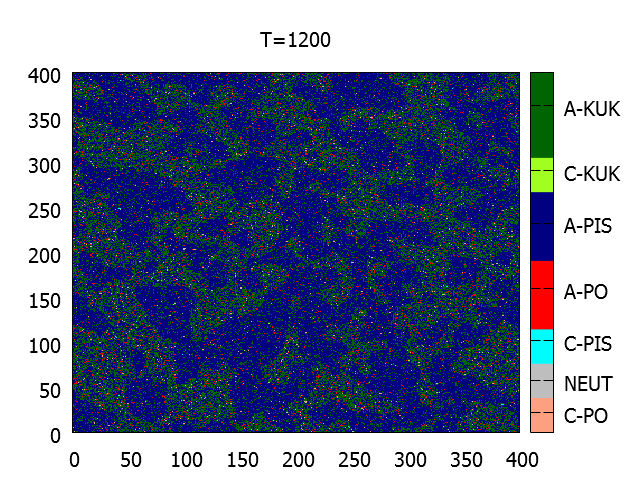} 
\caption{A snapshot of the system configuration at $T=1200$.  Kukiz support base has replaced most of the PO domains, in adddition,   PiS has gained significantly. The few remaining PO agents  are now in the agitated state. \label{fig:frame1200} }
\end{figure}

\begin{figure}
\includegraphics[scale=0.4]{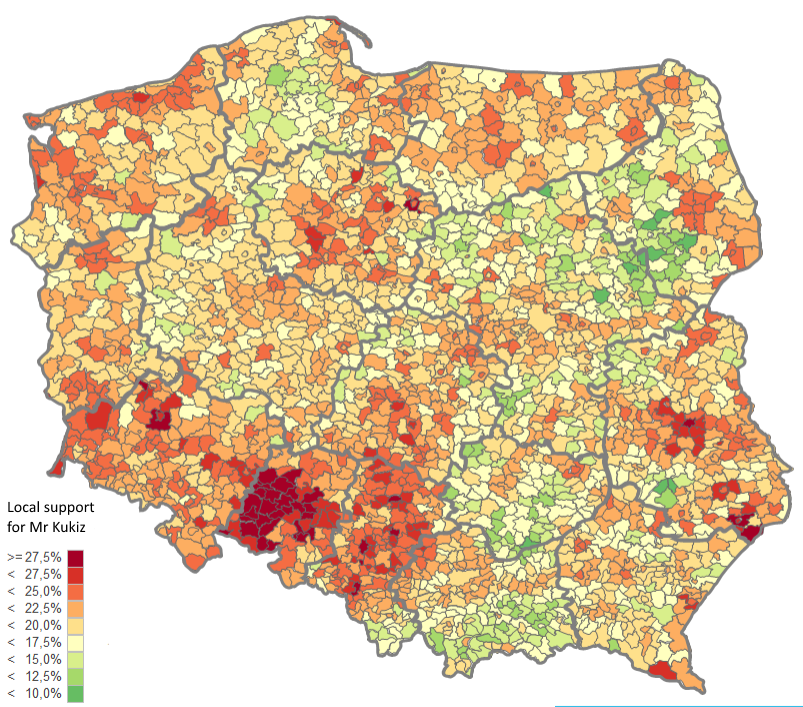} 
\caption{Geographical distribution of the support for Mr Kukiz in the first round of the presidential elections, May 2015. The highest support is located in some of the constituencies in which previously PO has enjoyed the highest popularity (see Figure\ref{fig.geography}). Source:  \url{http://skuteczneraporty.pl/blog/pawel-kukiz-rozgrzal-polske-na-kogo-zaglosuja-jego-zwolennicy-w-ii-turze/}.  \label{fig:kukizgeo} }
\end{figure}

It is quite interesting to compare the model process for the invasion of support for the newcoming party shown in Figures~\ref{fig:frame600} and \ref{fig:frame1200}  with the actual distribution of voter support for Mr Kukiz in the presidential elections, shown in Figure~\ref{fig:kukizgeo}. As expected, the highest support for Mr Kukiz occurred mostly in constituencies which were previously dominated by PO, while low support was correlated with the dominance of PiS. The correlation is not perfect -- but it should be noted that while in the model the domains in Figures~\ref{fig:frame199} and \ref{fig:frame600} contain almost only the supporters of a single party, the real constituencies shown in the maps of Figure~\ref{fig.geography} always contain some supporters of the local minority, so even in the PiS-preferring regions of Poland there are always some PO supporters, who may be the targets for Mr Kukiz movement.

\subsection{Effects of the change of newcomer strategy}
\label{sec:results}

The ``replacement" of PO by the Kukiz party shown in Figure~\ref{fig:evolution2} may be called an ideal strategy for the new movement. It is however, quite interesting, that due to lack of experience (and possibly other factors, such as the difference between the presidential campaign, focusing on a single personal image and the forthcoming parliamentary one, which requires organization and massive local presence), Mr Kukiz has changed his media strategy. There are many more messages that are perceived as irrational by the non-supporters, as well as a significant number of quarrels within the campaign staff, resulting in demobilizing internal messages as well.  The change has been incorporated into the model via parameter adjustment, shown in Table~\ref{tab:propaganda3}.

\begin{figure}
\includegraphics[scale=1.2]{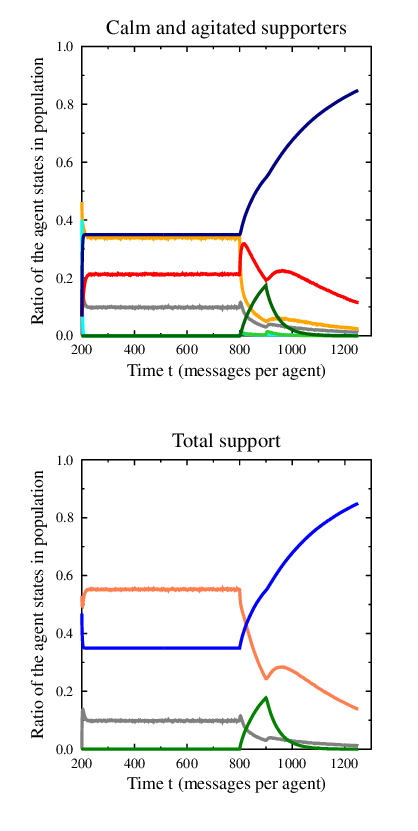} 
\caption{An example of the evolution of the support for the three political camps. At $T_3=900$ Kukiz party changes its communication strategy, and in doing so loses its momentum: instead of overpassing PO its support drops back to zero. Top panel: calm and agitated supporters; orange -- calm PO agents, red -- agitated PO agents; light blue -- calm PiS agents, dark blue -- agitated PiS agents; light green -- calm Kukiz agents, dark green -- agitated Kukiz agents; grey -- neutral agents. 
Bottom panel: evolution of the total number of supporters for each party. Orange -- PO, dark blue -- PiS, green -- Kukiz. \label{fig:evolution0}}
\end{figure}

\begin{figure}
\includegraphics[scale=0.4]{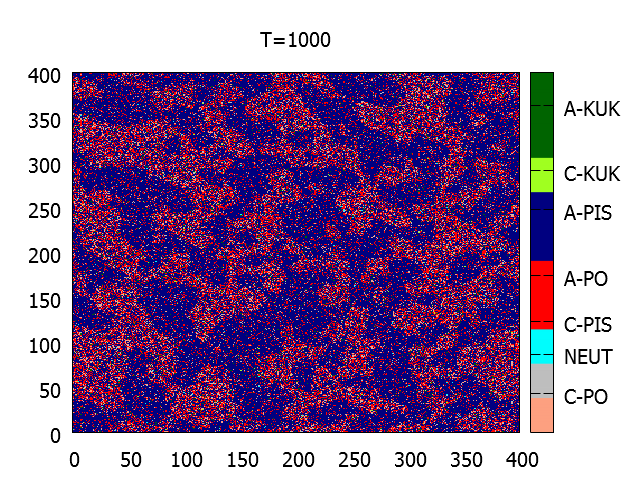} 
\caption{A snapshot of the system configuration at $T=1000$ in the situation when Kukiz party has changed the strategy away from optimal.  As a result its support has mostly fallen down.  PiS has gained significantly, while PO regained some presence, however, most of its supporters are now in the agitated state. \label{fig:frame1000} }
\end{figure}

The change in the Kukiz propaganda strategy (or maybe just its execution?) has disastrous effects to its popularity. Instead of capturing a large part of the previous PO support base, 
the support for Kukiz party stalled and begun to drop.
The new strategy allows PO to regain some foothold, as shown in recent polls, although at the ``cost" of the fact that almost all its supporters are now in the agitated state. The possibility of a dialogue between PiS (which now holds a large and growing majority) and PO (whose support diminishes (Figure~\ref{fig:evolution0}) is virtually impossible due to emotional state their supporters are in. 

By comparing Figures~\ref{fig:parties2015} and \ref{fig:evolution0} we could attempt (for the given set of simulation parameters) to roughly map the time events of the real world to the Monte Carlo timeline. The $T\sim 750$ would correspond to the beginning of 2015. $T=900$ would correspond to end of May, when Mr Kukiz popularity has reached its peak.
Thus 150 MC steps is about 5 months of real time. The date of writing this manuscript (July 10th) would correspond to $T=940$, well in agreement with the observed drop in the Kukiz poll figures and the small surge in PO support.  Within this assumptions, the parliamentary elections (scheduled for mid-October) would occur at $T=1050$, leaving Mr Kukiz little chances to gain significant number of seats. 
Of course, this prediction depends very much on the continuation of the current propaganda strategy. An example of such hypothetical scenario is presented in Section~\ref{sec:qualitative}.

\section{Discussion}
\label{sec:discussion}

\subsection{Model limitations}
\label{sec:limitations}

We are fully aware of severe limitations of our model, as an attempt to describe, even qualitatively, a real social situation. 

Firstly, the model does not take into account the appearance of new voters. Bearing in mind that the young voters are influenced to a very large extent by their peers (rather than by their elders),  their entry into the electoral system  disturbs it in more way than one. An extension of the current model aimed at allowing such dynamical `flow' of voters is planned.

Secondly, the voting preferences depend to a large extent on specific events, such as scandals involving politicians, or even results of football matches. These events typically have short term effects, but their accumulation may shift the balance more permanently.  Of course, the model does not contain such events and their effects. Moreover, the assumed asymmetry in the propaganda (only calm messages for PO, only aggressive messages from PiS) does not correspond to reality. In particular, the PO communications strategy contains some admixture of messages promoting fear of what would happen should PiS come into power. These messages, while fewer in number and visibility, were always present, and may serve to strengthen the resolve of the PO electoral base.

Thirdly, the model allows tremendous flexibility in the choice of the initial parameters for the simulations. It is very difficult to decide which set would correspond best to the observations. This limitation includes also the time dependence: there is no specific way to map the time measured in Monte Carlo steps  to years or weeks. 

Lastly, the model focuses on just three parties, while the actual situation is much more complex. As may be seen from Figure~\ref{fig:parties} until the end of 2014 the summed popularity of the ``minor" parties (TR, SLD, PSL) exceeded the difference between PO and PiS, so that these parties could aspire to tip the scales in the political situation -- a role quite successfully performed by PSL as PO coalition partner. Moreover, the model does not take into account the existence of inflexible supporters for the various parties, assuming that in principle all agents are free to change their sympathies. The introduction of small groups of die-hard supporters would change the quantitative results of the model.

\subsection{Qualitative discussion of possible future trends}
\label{sec:qualitative}

\begin{figure}
\includegraphics[scale=1.0]{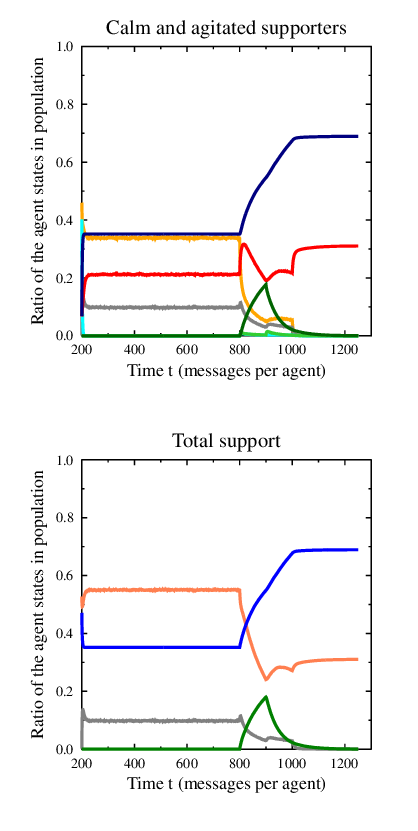} 
\caption{An example of the evolution of the support for the three political camps, when PO takes corrective action at $T_4=1000$. Top panel: calm and agitated supporters; orange -- calm PO agents, red -- agitated PO agents; light blue -- calm PiS agents, dark blue -- agitated PiS agents; light green -- calm Kukiz agents, dark green -- agitated Kukiz agents; grey -- neutral agents. 
Bottom panel: evolution of the total number of supporters for each party. Orange -- PO, dark blue -- PiS, green -- Kukiz. \label{fig:evolution1}}
\end{figure}

Despite the problems mentioned above, the model, through a simple action of taking into account the asymmetry  of communication strategies of the two dominant parties, reproduces the recent shift in the Polish politics surprisingly well.   

It is quite interesting to note that an additional support for the model may be derived from the fate of the Ruch Palikota/ Twoj Ruch (TR) party. 
This party was formed before the elections in 2011, by a prominent member of PO, Mr Janusz Palikot. It has very quickly gathered sizable support, as shown in Figure~\ref{fig:parties}. 
In the 2011 parliamentary elections it became the third power, surpassing both PSL and SLD, and gathering 10.02\% of the votes. 
These votes were mostly from previous supporters of PO. In fact the evolution of PO, PiS and TR support in 2011--2014 strongly resembles the shape shown by the model simulations. TR success is accompanied by the fall in PO popularity and growth in support of PiS. In fact, the geography of support for TR in the 2011 parliamentary elections follows the same pattern as the one mentioned in connection with Mr Kukiz in Figure~\ref{fig:kukizgeo}, that is the highest popularity was achieved in some regions previously dominated by PO, while the gains were the weakest in the regions dominated by PiS. The similarity between the two parties rests largely on their dependence on the strong personalities of their leaders. It is quite interesting that for both parties it is possible to use as the trends (in Figures~\ref{fig:parties} and \ref{fig:parties2015}) the same form of the fitting function
$A \exp(-(T-T_0)/\tau_2) / (1+\exp(-(T-T_0)/\tau_1)$, being a combination of logistic growth with the speed dictated by $\tau_1$ and exponential decay, with the speed dictated by $\tau_2$. The best fit values of these growth/decay times are, respectively $\tau_1=0.04$ year and $\tau_2=0.071$ year for Kukiz party (although the latter value is obtained from just a few data points), and  $\tau_1=0.081$ year and $\tau_2=1.83$ year for TR.

The lack of total fall of PO support due to the appearance of TR on the scene and the subsequent upturn of its rankings may be attributed to the shift in the party political strategy, taking a decidedly left turn, and to the personal success of Mr Tusk, in his election to the post of the President of the European Council. 
The success of the shift towards the left was made possible by totally chaotic management of the social-democratic party (SLD), which has lost almost 10\% points, mostly moving to the PO base. 
With respect to the current threat posed to PO by Mr Kukiz movement, we note that the scale of the intrusion is almost two times higher, and that there are no more such ``reserves" -- the minority parties are already reduced to the single figure die-hard core electorates, and thus there is no place left for another  policy shift for PO.

It may be worthwhile to discuss the long term trends leading to PiS overall domination, visible in all both Figures~\ref{fig:evolution2} and \ref{fig:evolution0}. They are due to total emotional mobilization of the PiS supporters thanks to the barrage of aggressive media messages. Still, should PiS win the upcoming elections (and thus have both the presidential power and the parliamentary and government one) at some point or another this success would have to be reflected in the communication strategy. After the initial period of blaming everything on the previous incompetent PO-led government, one would expect a growing admixture of self-promoting success stories. Thus a growing part  of the PiS electorate would become calm --- and therefore open to arguments by any upcoming party (for example social-democratic left, which was quite weak since 2007). 
On the other hand, it is possible to attain 90\% support or higher by relying on agitation rising propaganda. One way of maintaining this as a long term strategy is to shift the social attention and to focus it on external enemies. Such approach is being used, with huge success by the Russian leader, who, despite the increasingly bad economic situation, enjoys very strong and stable support (remaining steady at the 60\%--70\% range during the period between 2011 and 2014, and jumping up to over 80\% since the Russian invasion of Ukraine). We remind here that the focus on external enemies as a binding force is not a uniquely Russian recipe -- it has been used by  rulers in many countries and in many historical epochs. So, it could also offer a possibility as a long-term PiS strategy for total dominance.

The model allows us even to consider some speculations about the future strategies.  Suppose that the reaction to the PiS successes would be a complete change of communications strategy of PO --- moving away from the calming messages of `warm water in the tap' and `Europe's green island' and towards the scary stories of the horrors that the coming PiS led government will bring. Such shift is already beginning to occur in reality. These agitating messages would be addressed not at the PiS voters (whom it is impossible to convert) but at the PO own support base, turning it from the calm to the agitated state. If successful, such change would make the PO supporters immune to the PiS advances and create another stable equilibrium. Depending on the effectiveness of the campaign and on the time $T_4$ at which it is started, the resulting political landscape would contain a different ratio of PiS and PO supporters. An example of such evolution of the system is presented in Figure~\ref{fig:evolution1}. Propaganda ratios are in Table~\ref{tab:propaganda4}. 

While the solution described above preserves political plurality, it has a definite drawback: practically all agents in the system are in agitated states, unable do rationally process any arguments contrary to their current beliefs. The polarization becomes a permanent feature of the system, possibly leading to increased levels of misunderstanding, conflict and even violence. Such shift towards increased levels of animosity and polarization has been noted in American politics (see, for example, \cite{baum08-1,baldassari07-1,prior12-1}). Unfortunately, our model suggests that Poland is heading either  for  a situation in which the government is dominated by a single party (with an initially a strong, negative agenda), or for a system with increased polarization without hope for a consensus. Of course, real life may hold some surprises, and the predictions described above may be undermined by its weaknesses mentioned in section~\ref{sec:limitations}.

\appendix*
%
\section{List of propaganda ratios}

The values of the simulation parameters governing the propaganda streams of each party were chosen in a way to reproduce qualitatively the observations (up to July 10th, 2015). At the same time they correspond qualitatively to the ratios of various types of communications in the real world (as perceived by the author of the paper). 

\begin{table}[h]
\begin{tabular}{|p{4cm}|c|}
 \hline 
\textbf{ Type of message} & \textbf{Relative ratio} \\ 
 \hline 
 PO internal mobilizing & 0.0 \\ 
 \hline 
 PO internal demobilizing & 0.1 \\ 
 \hline 
 PO external rational & 0.25 \\ 
 \hline 
 PO external irrational & 0.0 \\ 
 \hline 
 \textbf{Total PO} & \textbf{0.35} \\ 
 \hline  
 PiS internal mobilizing & 0.4 \\ 
 \hline 
 PiS internal demobilizing & 0.0 \\ 
 \hline 
 PiS external rational & 0.0 \\ 
 \hline 
 PiS external irrational & 0.05 \\ 
 \hline 
 \textbf{Total PiS} & \textbf{0.45} \\ 
 \hline  
 \end{tabular}  
\caption{Propaganda types ratios for $T$ between $T_1=200$ and $T_2=800$.
\label{tab:propaganda1}}
\end{table}

\begin{table}[h]
\begin{tabular}{|p{4cm}|c|}
 \hline 
\textbf{ Type of message} & \textbf{Relative ratio} \\ 
 \hline 
 PO internal mobilizing & 0.0 \\ 
 \hline 
 PO internal demobilizing & 0.07 \\ 
 \hline 
 PO external rational & 0.18 \\ 
 \hline 
 PO external irrational & 0.0 \\ 
 \hline 
 \textbf{Total PO} & \textbf{0.25} \\ 
 \hline  
 PiS internal mobilizing & 0.25 \\ 
 \hline 
 PiS internal demobilizing & 0.0 \\ 
 \hline 
 PiS external rational & 0.05 \\ 
 \hline 
 PiS external irrational & 0.05 \\ 
 \hline 
 \textbf{Total PiS} & \textbf{0.35} \\ 
 \hline  
Kukiz internal mobilizing & 0.1 \\ 
 \hline 
Kukiz internal demobilizing & 0.0 \\ 
 \hline 
Kukiz external rational & 0.05 \\ 
 \hline 
Kukiz external irrational & 0.05 \\ 
 \hline 
 \textbf{Total Kukiz} & \textbf{0.20} \\ 
 \hline  
 \end{tabular}  
\caption{Propaganda types ratios for $T$ after the appearance of Kukiz party at $T_2=800$.
\label{tab:propaganda2}}
\end{table}

\begin{table}[h]
\begin{tabular}{|p{4cm}|c|}
 \hline 
\textbf{ Type of message} & \textbf{Relative ratio} \\ 
 \hline 
 PO internal mobilizing & 0.0 \\ 
 \hline 
 PO internal demobilizing & 0.07 \\ 
 \hline 
 PO external rational & 0.18 \\ 
 \hline 
 PO external irrational & 0.0 \\ 
 \hline 
 \textbf{Total PO} & \textbf{0.25} \\ 
 \hline  
 PiS internal mobilizing & 0.25 \\ 
 \hline 
 PiS internal demobilizing & 0.0 \\ 
 \hline 
 PiS external rational & 0.05 \\ 
 \hline 
 PiS external irrational & 0.05 \\ 
 \hline 
 \textbf{Total PiS} & \textbf{0.35} \\ 
 \hline  
Kukiz internal mobilizing & 0.05 \\ 
 \hline 
Kukiz internal demobilizing & 0.05 \\ 
 \hline 
Kukiz external rational & 0.0 \\ 
 \hline 
Kukiz external irrational & 0.10 \\ 
 \hline 
 \textbf{Total Kukiz} & \textbf{0.20} \\ 
 \hline  
 \end{tabular}  
\caption{Propaganda types ratios for $T$ after the change  of Kukiz party communication strategy at $T_3=900$, when more demobilizing and irrational messages appear in their propaganda.
\label{tab:propaganda3}}
\end{table}

\begin{table}[h]
\begin{tabular}{|p{4cm}|c|}
 \hline 
\textbf{ Type of message} & \textbf{Relative ratio} \\ 
 \hline 
 PO internal mobilizing & 0.1 \\ 
 \hline 
 PO internal demobilizing & 0.0 \\ 
 \hline 
 PO external rational & 0.15 \\ 
 \hline 
 PO external irrational & 0.0 \\ 
 \hline 
 \textbf{Total PO} & \textbf{0.25} \\ 
 \hline  
 PiS internal mobilizing & 0.25 \\ 
 \hline 
 PiS internal demobilizing & 0.0 \\ 
 \hline 
 PiS external rational & 0.05 \\ 
 \hline 
 PiS external irrational & 0.05 \\ 
 \hline 
 \textbf{Total PiS} & \textbf{0.35} \\ 
 \hline  
Kukiz internal mobilizing & 0.05 \\ 
 \hline 
Kukiz internal demobilizing & 0.05 \\ 
 \hline 
Kukiz external rational & 0.0 \\ 
 \hline 
Kukiz external irrational & 0.10 \\ 
 \hline 
 \textbf{Total Kukiz} & \textbf{0.20} \\ 
 \hline  
 \end{tabular}  
\caption{Propaganda types ratios for $T$ after the hypothetical change  of PO communication strategy at $T_4=1000$, when the demobilizing messages are dropped out and replaced by mobilizing ones.
\label{tab:propaganda4}}
\end{table}

\end{document}